\documentclass[showpacs,preprintnumbers,amssymb,pla]{revtex4}
\usepackage{graphicx}
\usepackage{color}
\usepackage{dcolumn}
\usepackage{bm}
\usepackage{longtable}
\usepackage{graphics}
\date{\today}
\begin{document}

\title{Testing non-locality of single photons using cavities}

\author{T. Pramanik\footnote{tanu.pram99@bose.res.in}$^1$, S. Adhikari\footnote{satyabrata@iiopb.res.in}$^2$, A. S. Majumdar\footnote{archan@bose.res.in}$^3$, D. Home\footnote{dhome@bosemain.boseinst.ac.in}$^4$}

\address{$^{1,3}$ S. N. Bose National Centre for Basic Sciences,
Salt Lake, Kolkata 700 098, India}

\address{$^2$ Institute of Physics, Sainik School Post, Bhubaneswar 751005, India}

\address{$^4$ CAPSS, Department of Physics, Bose Institute, Kolkata 700091, India}

\begin{abstract}
A scheme is formulated for testing nonlocality of single photons by considering
the state of a single photon that could be located within one of two spatially
separated cavities. The outcomes of four experiments on this state involving
the resonant interactions of two-level atoms with these cavities and a couple of
other auxiliary ones is shown to lead to a contradiction with the criterion
of locality. 
\end{abstract}


\maketitle

\paragraph{Introduction.\textemdash} 

In classical physics two distantly separated particles obey Einstein's 
locality, i.e., the outcome of a measurement 
on one of the particles does not effect what is being measured on the other. 
On the other hand, the nonlocal nature of the quantum world has for
long attracted deep attention among physicists ever since the
famous EPR paper \cite{EPR}. Specifically, though information cannot be 
transmitted between
spacelike separated observers, any realistic hidden variable theory
capable of reproducing the results of quantum theory would have to be 
nonlocal. A precise identification of nonlocality as a
crucial ingredient of distinction between classical and quantum physics was
performed by Bell in his seminal paper \cite{Bell}.  The violation of 
Bell's inequality which has been demonstrated  in a number of experiments 
\cite{Expt1}, 
underlines the nonlocality of quantum states of two or more particles.

If nonlocality is to be regarded as an inherent feature of the quantum
world, it is difficult to understand why this feature should be manifest only
at the level of two or more particles. A quantum state
should reveal its nonlocal property irrespective of the number of particles
associated with it. At the field theoretic level particles are regarded as
excitations of the quantum vacuum, and there is no fundamental difference 
between a single particle excitation, and a two-particle one. Specifically, 
a single particle
should also be able to exhibit nonlocality under particular circumstances, as
was indeed indicated by Einstein in 1927 \cite{solvay} while presenting 
the collapse of
a single particle wave packet to a near eigenstate as an example of quantum
nonlocality. It took more than half a century since then for the concept of 
nonlocality of single particles to be more precisely formulated by  Tan, Walls 
and Collett \cite{TWC} through the idea that measurements made
on two output channels from a source could violate locality even if one
particle is emitted from the source at a time. However, additional assumptions
in this proposal narrowed down the scope of its implementation 
considerably \cite{twc2}. In this context it may be noted that a single
particle could carry two or more degrees of freedom (i.e., the spatial
and polarization variables in case of a photon), and it is possible to
demonstrate the violation of Bell-type inequalities in such cases signifying
the violation of non-contextuality at the level of a single 
particle \cite{home}. Further, entanglement between different degrees of
freedom of the same particle could be exploited as resource for performing
information processing tasks \cite{pramanik}. On the other hand, the 
establishment of nonlocality at the level of a single particle has
still remained a subtle issue.

A demonstration of quantum nonlocality for two entangled particles
without the use of mathematical inequalities, was formulated
by Hardy  \cite{hardy0}.  Subsequently, he proposed a scheme to demonstrate
the nonlocality of single photons \cite{Hardy} without the 
supplementary assumptions 
made in the work of Tan, Walls and Collett \cite{TWC}. Hardy's scheme was
criticized for not being experimentally realizable by Greenberger, Horne
and Zeilinger (GHZ) \cite{ghz} who in turn proposed their 
own scheme which required
additional particles for implementation. Here again, the issue of whether
nonlocality is purely a multipartite effect could not be settled since it 
could be debated that the additional particles were responsible for
introducing nonlocality into the system. Another proposal showing an 
incompatibility between quantum mechanics and any local deterministic 
ontological model in terms of particle coordinates was formulated
for single photon states \cite{HA}. Recently, Dunningham and Vedral 
\cite{DV} have formulated a scheme for demonstrating nonlocality of single
photons, overcoming several problems of earlier proposals. 
Their scheme which
relies on the use of mixed states, is experimentally realizable, as claimed
by the authors, but has still not been practically performed. 

The notion of single particle nonlocality is  of
such conceptual importance in the physical interpretation of quantum
theory, that it is worthwhile to think of more proposals in order to 
demonstrate it beyond any reasonable doubt. In this regard, it may be noted 
that as different from the case of nonlocality of two particles,  single 
particle nonlocality has still not been conclusively demonstrated 
experimentally in a
manner free of conceptual loopholes. Note here that an experiment performed 
earlier by Hessmo et al. \cite{hessmo} for exhibiting single photon 
nonlocality was
based on the schemes of Tan, Walls and Collett \cite{TWC} and Hardy 
\cite{Hardy}, and hence, not free of the conceptual problems raised by
GHZ \cite{ghz} and others \cite{DV}. The scheme proposed by Dunningham and
Vedral \cite{DV} promises to circumvent those problems, but is yet to be 
experimentally implemented.  With the above
perspective, in this paper we present a proposal for demonstrating nonlocality
of single photons inside cavities. The formulation of our scheme is based on 
atom-photon interactions in cavities, a well-studied arena on which controlled 
experiments have been performed for many years now \cite{Haroche}.   
The ingredients for our proposal are two-level atoms, and single-mode high-Q
cavities which are tuned to resonant transitions between the atomic levels. For 
example, the use of Rydberg atoms and microwave cavities in testing several
fundamental aspects of quantum mechanics have been proposed \cite{majumdar},
and various interesting experiments have been performed by keeping dissipative
effects under control \cite{cavexpts}. 

\paragraph{Atom-cavity interaction dynamics.\textemdash}

We consider the dynamics of a two-level atom passing through a cavity, which
under the dipole and rotating wave approximations is described by the 
Jaynes-Cummings interaction
Hamiltonian \cite{apr2,apr4}
\begin{eqnarray}
H^I(r)= G(r) (a \sigma_{+}+a^{\dagger} \sigma_{-}),
\end{eqnarray}
where $\sigma_{\pm}=\sigma_x\pm\sigma_y$ denote the Pauli spin 
operators for a two-level atom, and $a$, $a^{\dagger}$ are the annihilation and 
creation operators, respectively,  for a 
photon in single mode cavity. The atom-field 
coupling strength may be expressed as  
$G(r)=\Omega_0(\hat{\bf{d}}_{eg}. \hat{\bf{e}}(r))f(r)$, where $\Omega_0$ is 
the peak atomic Rabi frequency, $\hat{\bf{d}}_{eg}$ is the orientation of the 
atomic dipole moment, and $\hat{\bf{e}}(r)$ is the direction of the electric 
field vector at the position of the atom. The 
profile $f(r)$  has an exponential envelop centered about the point in 
the atom's trajectory that is nearest to the centre of the cavity, $r_0$
\cite{apr2,apr4}. 
Within this envelope, the field intensity oscillates sinusoidally, and for 
the fixed dipole orientation, variations in the relative orientation of the 
dipole and electric field gives a sinusoidal contribution, i.e.,
\begin{eqnarray}
f(r)=e^{-\frac{|r-r_0|}{R_{def}}} \cos[\frac{\pi}{a_l}(r-r_0)],
\end{eqnarray}
where $R_{def}$ defines the spatial extent of the mode which is at most a few 
times the lattice constant ($a_l$) for a strongly confined mode in a 
photonic band gap \cite{apr2}.

The atom-field state after an initially excited atom has passed through a 
cavity (which is initaially is in zero photon state) can be written as 
\begin{eqnarray}
|\phi_e\rangle=\alpha_1(t) |e\rangle |0\rangle + \alpha_2(t) |g\rangle |1\rangle,
\end{eqnarray}
where $\alpha_{1(2)}$ is the amplitude of the atom being in excited (ground) 
state and $t$ is the interaction time of the atom with the cavity, with
\begin{eqnarray}
\alpha_1(t)=\cos[\int_0^t G(t^{\prime}) dt^{\prime}] \nonumber \\
\alpha_2(t)=\sin[\int_0^t G(t^{\prime}) dt^{\prime}],
\end{eqnarray}
where we have replaced $(r-r_0)$ by $(vt-b)$ with $v$ being the velocity of 
the atom in the cavity and $2b$ the effective length of interaction in the 
cavity. Setting the value of the interaction time $t = 2b/v$, it is possible to
obtain the exact expressions for the $\alpha_{1(2)}$, i.e., 
\begin{eqnarray}
\alpha_{1(2)} = \cos(\sin)\Biggl[\frac{2 a_l R_{def} \Omega_0 k e^{-\frac{b}{R_{def}}} \left(a_l e^{\frac{b}{R_{def}}}+\pi  R_{def} \sin \left(\frac{\pi  b}{a_l}\right)-a_l \cos \left(\frac{\pi  b}{a_l}\right)\right)}{v
   \left(a_l^2+\pi ^2 R^2_{def}\right)}\Biggr]
\end{eqnarray}
with $k = \hat{{\bf d}}_{eg}. \hat{{\bf e}}({\bf r})$, which can be chosen
to take values from $0$ to $1$. Following \cite{apr4},   
we henceforth set $a=624 nm = R_{def}$,   $b=10 R_{def}$, and 
$\Omega_0 = 1.1 \times 10^{10} rad/s$ in our calculations.

\paragraph{State preparation.\textemdash} 

The scheme for demonstrating nonlocality that we use in the present work
is set up as follows.  Let Alice and Bob be two  spatially well separated
parties who possess
two cavities  $C_1$ and $C_2$, respectively, an atom $a_1$ and a 
detector $D_1$ used
in the state preparation process. Alice
and Bob also possess an auxiliary cavity and an additional atom each, 
viz., $C_4$, $a_3$ and $C_3$, $a_2$ respectively, as
well as two detectors $D_2$ and $D_3$, respectively, which they could either
use or remove depending on the choice of the four different experiments
they perform, as described below.


\begin{figure}[h]
{{\resizebox{8.0cm}{10.0cm}{\includegraphics{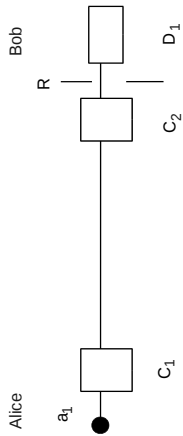}}}}
\caption{\footnotesize Experimental set up for creating the state given by Eq.(\ref{atom1g}). A two-level atom in its upper level traverses the cavities $C_1$ and $C_2$, before being detected in its lower level by the detector $D_1$.
($v_{a_1} = 179 \text{m/s}$, $k(a_1;C_1) = 0.979$, $k(a_1;C_2) = 1$).}
\end{figure}

We begin by describing the state preparation process for single photons
(see Fig. 1),
that we will use for the argument of nonlocality.  
Initially, the two cavities $C_1$ and $C_2$ are empty, i.e., in zero photon 
states denoted by $|0\rangle_{C_1}$ and  
$|0\rangle_{C_2}$. Now an atom (say atom-1), initially in the upper of its two
possible levels, $|e\rangle_{a_1}$ traverses $C_1$ with flight time 
$t_{11}$ and then traverses $C_{2}$ with flight time $t_{12}$. The cavities
are tuned to the resonant frequency for transitions between the upper and 
lower levels of the atom.
Now,  a $\frac{\pi}{2} R_1$ Ramsey pulse is applied at a frequency resonant 
with the atomic transition  at the departure path of atom-1
(the $R_1$ pulse is essentially a part of the
detection mechanism \cite{Haroche} and causes the transitions
$|e\rangle \to \frac{|e\rangle-|g\rangle}{\sqrt{2}}$ and $|g\rangle \to \frac{|e\rangle+|g\rangle}{\sqrt{2}}$).
Let us consider the case when  atom-1 is detected in the state
 $|g\rangle_{a_1}$ at the detector $D_1$. Since the atom is intially prepared in
its excited state, and both the cavities are devoid of any photons intially,
the atom can make a transition to its lower state only by dumping a single
photon in either of the two cavities. It then follows that after detection of
the atom the state of the 
single photon  is given by
\begin{eqnarray}
|\psi \rangle = - \alpha_1(t_{11}) \alpha_1(t_{12}) |0\rangle_{C_1}|0\rangle_{C_2} 
+ \alpha_2(t_{11})|1\rangle_{C_1}|0\rangle_{C_2} 
+ \alpha_1(t_{11}) \alpha_2(t_{12}) |0\rangle_{C_1}|1\rangle_{C_2} 
\label{atom1g}
\end{eqnarray}
where  the second (third) term
on the r.h.s represents the single photon in cavity $C_1$ ($C_2$) and no 
photon in cavity $C_2$ ($C_1$). The first term arises as a result of the  
$\frac{\pi}{2} R_1$ pulse introduced as part of the detection mechanism.
This completes our state preparation process. Note that though 
Eq.(\ref{atom1g}) representing the state prepared for the following 
experiments is similar to the single photon states used by Hardy \cite{Hardy} 
and Dunningham and Vedral \cite{DV} in their arguments on single photon 
nonlocality, the physical constituents are quite different.  

It is important to mention here that in the present scheme 
the state preparation process is separate from the experimental procedure
(described below) to infer the nonlocality of the prepared state.
Any additional particles introduced by Alice 
and Bob in the experiments using the state (\ref{atom1g}), will not cause
any additional nonlocality to be introduced in the state (\ref{atom1g}),
as is evident from the following argument.
The combined state of all the resources possessed by Alice and Bob together
at the beginning of the experiment can be described as
\begin{eqnarray}
|\psi\rangle_{expt} = |\psi \rangle \otimes |0\rangle_{C_4} 
\otimes |e\rangle_{a_3} \otimes |0\rangle_{C_3} \otimes |e\rangle_{a_2} 
\label{expt}
\end{eqnarray}
where the second and fourth term on the r.h.s correspond to the zero
photon states inside auxiliary cavities possessed by Alice and Bob 
respectively, and
the third and fith term correspond to the excited atomic states of
Alice and Bob, respectively. In their experiments Alice and Bob have
the choice of performing local unitary operations using their above
resources, and detecting their atoms by their respective detectors $D_2$ and
$D_3$. It is clear that such operations will not in any way impact the
nonlocal property of the state  $|\psi \rangle$ (\ref{atom1g}) that we
wish to demonstrate. Any nonlocal feature must already have been 
introduced at the stage of preparation of the state described above.

\paragraph{The scheme.\textemdash}

Alice and Bob have two options each to proceed. 
In one of them, they can find out
directly whether the photon is inside their own cavity. Operationally,
Alice (Bob) has to take an auxiliary atom in a ground state and pass it 
through her (his)  cavity  with the choice of parameters $v = 161 m/s$ and
$k=1$, (making $\alpha_1 = 0$ and $\alpha_2 = 1$), ensuring that if the 
photon is present inside the cavity, then
the atom is detected in the state  $|e\rangle$. On the other hand, if the atom 
is detected in the state   $|g\rangle$, then Alice (Bob) concludes that the
photon was not present in her (his) cavity. In the second option Alice (Bob)
places an auxiliary (initially empty) cavity in  front her (his) cavity. 
Then an atom in the state $|e\rangle$ is sent through the two cavities 
successively before being detected by $D_2$ ($D_3$). These choices lead us to 
the following four experiments.

{\it Experiment 1.-} Alice and Bob both decide to check whether the photon is
present inside their own cavity, $C_1$ and $C_2$, respectively. In this case, 
it is clear that (from Eq.\ref{atom1g}) Alice (Bob) either finds a single
photon inside her (his) cavity, or nothing. They can not both find one 
photon each
in their cavities, as the atom-1 dumps only one photon either in $C_1$ or 
in $C_2$. This means that detecting one photon by Alice and one photon by Bob 
never happens together.

\begin{figure}[h]
{{\resizebox{8.0cm}{10.0cm}{\includegraphics{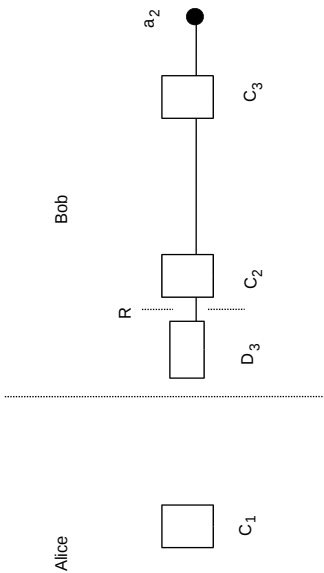}}}}
\caption{\footnotesize Set-up for Experiment-2. Alice checks directly for
a photon in cavity $C_1$. Bob passes the atom $a_2$, initially in its upper
level, through $C_3$ and $C_2$, before detecting it at $D_3$ again in its
upper level. ($v_{a_2} = 161 \text{m/s}$, $k(a_2;C_2) = 1$, $k(a_2;C_3) <1$).}
\end{figure}

{\it Experiment 2.-} In this case Alice checks whether the photon is present
inside her cavity $C_1$, as in experiment 1. Bob takes his auxiliary cavity 
$C_3$ in state 
$|0\rangle_{C_3}$ and passes another atom $a_2$ in state $|e\rangle_{a_2}$ (shown
in Fig. 2) through $C_3$ and $C_2$ with flight times $t_{23}$ and $t_{22}$ 
respectively. Bob then applies a $\frac{\pi}{2} R_1$ pulse on $a_2$, before
detecting it at $D_3$.
Now consider the case when Alice detects no photon inside her cavity $C_1$. 
The probability for this case happening is given by  $\alpha_1^2(t_{11})$.
Then, after Bob passes $a_2$ through his cavities $C_3$ and $C_2$, and applies
the  $\frac{\pi}{2} R_1$ pulse, suppose that $a_2$ is detected in the state
 $|e\rangle_{a_2}$.  The atom $a_2$ was initially in the state  $|e\rangle_{a_2}$
as well, and hence this means that in its flight it has not lost its energy
by dumping any photon. Now choosing 
the velocity of $a_2$ (corresponding to the flight time $t_{22}$) through the
cavity $C_2$ to be $v_{a_2} = 161 m/s$ and $k(a_2;C_2)=1$,
it follows  that in this case
Bob's  state is given by
\begin{eqnarray}
|\psi \rangle_B = 
N_1((-\alpha_1(t_{12})\alpha_1(t_{23})+\alpha_2 (t_{12})\alpha_1(t_{23}))|1\rangle_{C_2}|0\rangle_{C_3}
 - (\alpha_1(t_{12})\alpha_2(t_{23})+\alpha_2(t_{12})\alpha_2(t_{23}))|0\rangle_{C_2}|1\rangle_{C_3})
\label{expt2R}
\end{eqnarray}
where, $N_1=1/\sqrt{1-2\alpha_2(t_{12})\alpha_1(t_{12})(\alpha_1^2(t_{23})-\alpha_2^2(t_{23}))}$.
Note here that with the above choice of the velocity 
$v_{a_2}$ and $k(a_2;C_2)$,  the possibility of obtaining a two-photon 
state such as 
$|1\rangle_{C_2}|1\rangle_{C_3}$ gets ruled out (since $\alpha_1(t_{22}) =0$
and $\alpha_2(t_{22}) = 1$). 
 Further, let us choose
$v_{a_1} = 179 \text{ m/s}$ and $k(a_1;C_2) = 1$ 
(i.e., $\alpha_1(t_{12}) = 1/\sqrt{2} = \alpha_2(t_{12})$), 
such that the first term on the r.h.s. of Eq.(\ref{expt2R}) vanishes. With
such a choice of the interaction 
parameters, it follows that Bob 
can find the photon in cavity $C_3$ only,
but not in $C_2$. In other words, in the case when Alice detects no photon
in her cavity $C_1$, if Bob detects a 
single photon, it must be in cavity $C_3$, and not in cavity $C_2$. Now,
reversing this argument, if Bob detects a photon in cavity $C_2$,
and nothing in $C_3$ (it follows from Eqs.(\ref{atom1g})
and (\ref{expt2R}), that such an outcome occurs with a finite probability
given by $1 - \alpha_1^2(t_{11}) \alpha_2^2(t_{23})$), then Alice 
{\it cannot} detect no photons inside her
cavity $C_1$, i.e., she must detect a single photon there, since this is 
the only other possible outcome.  

Note here that in the above argument we have specifically
chosen to describe the case when the atom is detected in the excited state 
 $|e\rangle_{a_2}$.
However, it needs to be mentioned that a similar argument can also be
constructed in the case when the atom is detected in the ground state 
 $|g\rangle_{a_2}$.
The only difference between the two cases are the values of the experimental
parameters required for the scheme to work out. For example, the state
corresponding to Eq.(8)  in the latter case would be given by
$N_1 [- \alpha_1(t_{23}) (\alpha_1(t_{12})+\alpha_2(t_{12}) |10\rangle_{C_2C_3} + \alpha_2(t_{23}) (\alpha_2(t_{12})-\alpha_1(t_{12})) |01\rangle_{C_2C_3}]$, where
$N_1^2=1/(1+2 \alpha_1(t_{12}) \alpha_2(t_{12}) (\alpha_1^2(t_{23})-\alpha_2^2(t_{23})))$ with the probability of detection of $a_2$ in ground state being
 $1/(2 N_1^2)$. It follows that in this case 
 one would require $v_{a_1} = 146 m/s$ (in stead
of $179 m/s$ as required in the former case) in order to ensure that the photon
is found in cavity $C_3$ and not in $C_2$.      

\begin{figure}[h]
{{\resizebox{8.0cm}{10.0cm}{\includegraphics{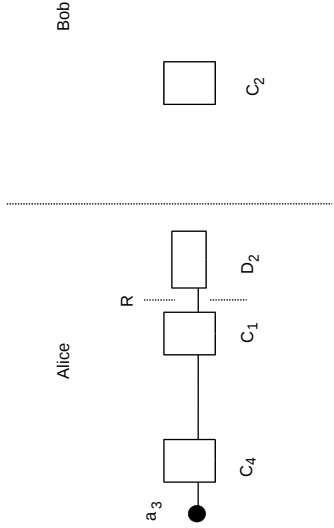}}}}
\caption{\footnotesize Set-up for Experiment-3. Bob checks directly for
a photon in cavity $C_2$. Alice passes the atom $a_3$, initially in its upper
level, through $C_4$ and $C_1$, before detecting it at $D_2$ again in its
upper level. ($v_{a_3} = 161 \text{m/s}$, $k(a_3;C_1) = 1$, $k(a_3;C_4) <1$).}
\end{figure}

{\it Experiment 3.-} This is similar to experiment-2, with the roles of Alice 
and Bob reversed.  Bob checks whether the photon is present
inside his cavity $C_2$, as in experiment 1. Alice takes her auxiliary cavity 
$C_4$ in state 
$|0\rangle_{C_4}$ and passes another atom $a_3$ in state $|e\rangle_{a_3}$ (shown
in Fig. 3) through $C_4$ and $C_1$ with flight times $t_{34}$ and $t_{31}$ 
respectively. Alice then applies a $\frac{\pi}{2} R_1$ pulse on $a_3$, before
detecting it at $D_2$.
Now consider the case when Bob detects no photon inside his cavity $C_2$.
Further, suppose that $a_3$ is detected by Alice in the state
 $|e\rangle_{a_3}$. Then, by choosing the value
$v_{a_3} = 161 \text{ m/s}$ and $k(a_3;C_1) = 1$, 
it follows that in this case Alice's  state is given by
\begin{eqnarray}
 |\psi\rangle_A =
N_3((- \alpha_1(t_{11})\alpha_1(t_{12})\alpha_1(t_{34})
+ \alpha_1(t_{34})\alpha_2(t_{11}))|0\rangle_{C_4}|1\rangle_{C_1} \nonumber\\
- (\alpha_1(t_{11})\alpha_1(t_{12})\alpha_2(t_{34}) 
+ \alpha_2(t_{34})\alpha_2(t_{11}))|1\rangle_{C_4}|0\rangle_{C_1})
\label{exp3R}
\end{eqnarray}
with  $N_3=1/\sqrt{\frac{\alpha_1^2(t_{11})}{2}+\alpha_2^2(t_{11})-\sqrt{2} \alpha_1(t_{11}) \alpha_2(t_{11}) (\alpha_1^2(t_{34})-\alpha_2^2(t_{34}))}$. Using the 
value of   $\alpha_1(t_{12}) = 1/\sqrt{2}$  from {\it Experiment-2},
it follows that the first term on the r.h.s. of Eq.(\ref{exp3R})
vanishes when we set
$v_{a_1}=179 \text{ m/s}$ and $k(a_1;C_1) =0.979$, 
(i.e., $\alpha_1(t_{11}) = \sqrt{2}\alpha_2(t_{11})$).
Hence, Alice can find the photon in cavity $C_4$ only,
but not in $C_1$. In other words, in the case when Bob detects no photon
in his cavity $C_2$, if Alice detects a 
single photon, it must be in cavity $C_4$, and not in cavity $C_1$. Now,
reversing this argument, if Alice detects a photon in cavity $C_1$,
and nothing in $C_4$ (it follows from Eqs.(\ref{atom1g})
and (\ref{exp3R}), that such an outcome occurs with a finite probability
given by  $1 -\alpha_2^2(t_{34})\alpha_1^2(t_{11})$), 
then Bob {\it cannot} detect no photons inside his
cavity $C_2$, i.e., he must detect a single photon there, since this is 
the only other possible outcome.   

\begin{figure}[h]
{{\resizebox{8.0cm}{10.0cm}{\includegraphics{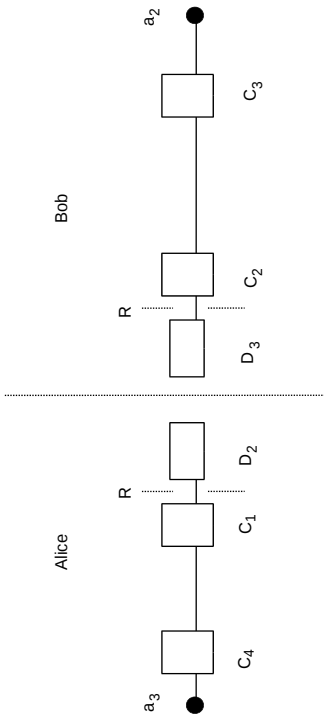}}}}
\caption{\footnotesize Set-up for Experiment-4. Alice does the same as she
did in experiment-3, while Bob does the same as in experiment-2.}
\end{figure}

{\it Experiment 4.-} In this experiment (shown in Fig. 4) Alice and Bob both 
use there auxiliary cavities. Alice passes her atom $a_3$ through $C_4$ and
$C_1$, and Bob passes his atom $a_2$ through $C_3$ and $C_2$. Further, both
apply $\frac{\pi}{2} R_1$ pulses on their atoms which are subsequently
detected in their upper states $|e\rangle_{a_3}$ and $|e\rangle_{a_2}$ 
respectively. One of the possibilities of this experiment is that Alice
detects a photon in cavity $C_1$ and nothing in $C_4$, while Bob detects a
photon in cavity $C_2$ and nothing in $C_3$, as is reflected from the
following term
\begin{eqnarray}
|\psi\rangle_{AB} = [- \alpha_1(t_{11})\alpha_1(t_{12})\alpha_1(t_{23})\alpha_1(t_{34})\alpha_2(t_{31})\alpha_2(t_{22}) 
+\alpha_1(t_{11})\alpha_1(t_{12})\alpha_1(t_{23})\alpha_1(t_{34})\alpha_2(t_{31}) \nonumber\\
+\alpha_2(t_{11})\alpha_1(t_{23})\alpha_1(t_{34})\alpha_2(t_{22})]
|1\rangle_{C_1}|1\rangle_{C_2}|0\rangle_{C_3}|0\rangle_{C_4} + ......
\end{eqnarray}
in their joint state $|\psi\rangle_{AB}$. Such an outcome occurs 
with the probability
$0.0847 \alpha_1^2(t_{23})\alpha_1^2(t_{34})$. Note here that one can
choose values for the parameters $k(a_2:C_3)$ and $k(a_3;C_4)$ such
that this probability is non-vanishing. The maximum probability $0.0847$ 
occurs for $k(a_2:C_3)=k(a_3;C_4)=0.8$ ($\alpha_1(t_{23})=\alpha_1(t_{34})=1$).

The result of experiment-4 leads to a contradiction when combined with the
other experiments, as follows. Once Alice finds a photon in cavity $C_1$,
following the logic of experiment-3 she infers that Bob must find a photon
if he were to check {\it directly} for it in his cavity $C_2$ {\it without} 
using auxiliary resources of $C_3$ and $a_2$. Similarly, on finding a photon
in cavity $C_2$, Bob infers using the logic of experiment-2 that Alice would 
find a photon in $C_1$ if she were to check {\it directly} for it in her 
cavity $C_1$ {\it without} using auxiliary resources of $C_4$ and $a_3$. 
However, they both cannot be right, since
it follows from the result of experiment-1 that both Alice and Bob could
never detect a photon each by {\it directly} checking for it in their
respective cavities $C_1$ and $C_2$ {\it without} using their auxiliary
resources. The  contradiction arises from the fact that the above inferences
of Alice and Bob are based on the criterion of locality \cite{Hardy}.
The consideration of locality leads to the assumption that the probability of 
Bob obtaining an outcome is independent of the experiment Alice performs, 
and vice-versa. There is no contradiction if one does not use this
assumption of locality, and hence, the conclusion follows about the
nonlocality of the single photon state (\ref{atom1g}).

\paragraph{Concluding remarks.\textemdash}

Before concluding, it is worth mentioning certain points of comparison of
our proposal for testing the nonlocality of single photon states using
cavities and two-level atoms, with the earlier schemes of Hardy \cite{Hardy}
and Dunningham and Vedral \cite{DV}. Apart from the analogous nature of
the argument leading to the above-mentioned contradiction with the locality
assumption, the algebra of the relevant states bears formal resemblance
to those used in the earlier works \cite{Hardy, DV}. This is to be expected
since at the level of state preparation what we have done in the present scheme 
is to replace the beam-splitter and incident vacuum modes by a two-level
atom passing through two intially empty cavities.  There are some additional
differences from the earlier schemes in the detection mechanism used
in the experiments-2, 3 and 4. Here we employ auxiliary cavities and additional
two-level atoms, in stead of the homodyne detection scheme. In the present 
scheme  two-photon terms simply drop out by the choice of interaction 
parameters, whereas, in the scheme \cite{DV} state
truncation is required to ensure that the possibility of the presence of 
two photon states is avoided. 
Note that the choice of the velocities of the atoms that we have proposed
in the various experiments ($v_{a_1} = 179 \text{m/s}$, and $v_{a_2} = 161 \text{m/s} = v_{a_3}$) fall within the thermally accessible range of velocities 
\cite{apr2,apr4}. The values for the other interaction parameter
($k\equiv  \hat{{\bf d}}_{eg}. \hat{{\bf e}}({\bf r})$) that we have chosen
($k(a_1;C_1) = 0.979$, $k(a_1;C_2) = k(a_2;C_2) = k(a_3;C_1) = 1$, and
$k(a_2;C_3) < 1$, $k(a_3;C_4) < 1$), should also be attainable. 
 Further, making use of resonant interactions
between atoms and cavities enables us to avoid using coherent \cite{Hardy}
or mixed \cite{DV} states that may be difficult to create experimentally.
To summarize, our proposal is based on generating atom-cavity entanglement
that has already been practically operational for many 
years now \cite{cavexpts}. Thus, our scheme
should facilitate testing the nonlocality of single photons
in an actual experiment free of conceptual loopholes.

{\it  Acknowledgements} ASM and DH acknowledge support from the DST project
no. SR/S2/PU-16/2007. DH thanks Centre for Science, Kolkata for support.

\pagebreak

\end{document}